\documentclass[12pt,preprint]{aastex}

\shorttitle{Consistent non-equilibrium dust cloud  formation}
\shortauthors{Helling, Dehn et al.}

\begin{document}

%% LaTeX will automatically break titles if they run longer than
%% one line. However, you may use \\ to force a line break if
%% you desire.

\title{Consistent simulations of substellar atmospheres\\ 
 and non-equilibrium dust-cloud  formation}

%% Use \author, \affil, and the \and command to format
%% author and affiliation information.
%% Note that \email has replaced the old \authoremail command
%% from AASTeX v4.0. You can use \email to mark an email address
%% anywhere in the paper, not just in the front matter.
%% As in the title, use \\ to force line breaks.

\author{Christiane Helling}
\affil{SUPA, School of Physics \& Astronomy, Univ. of St Andrews, North Haugh, St Andrews,  KY16 9SS, UK}
\email{Christiane.Helling@st-and.ac.uk}

\author{Matthias  Dehn}
\affil{Hamburger Sternwarte, Gojenbergsweg 112, 21029 Hamburg, Germany}

\author{Peter Woitke}
\affil{UK ATC, Royal Observatory, Blackford Hill, Edinburgh EH9 3HJ, UK}

\and

\author{Peter H. Hauschildt}
\affil{Hamburger Sternwarte, Gojenbergsweg 112, 21029 Hamburg, Germany}

%% Notice that each of these authors has alternate affiliations, which
%% are identified by the \altaffilmark after each name.  Specify alternate
%% affiliation information with \altaffiltext, with one command per each
%% affiliation.

%% Mark off your abstract in the ``abstract'' environment. In the manuscript
%% style, abstract will output a Received/Accepted line after the
%% title and affiliation information. No date will appear since the author
%% does not have this information. The dates will be filled in by the
%% editorial office after submission.

\begin{abstract}
We aim to understand cloud formation in substellar objects.  We
combined the non-equilibrium, stationary cloud model of Helling,
Woitke \& Thi (2008; seed formation, growth, evaporation,
gravitational settling, element conservation) with the general-purpose
model atmosphere code {{\sc Phoenix}} (radiative transfer, hydrostatic
equilibrium, mixing length theory, chemical equilibrium) in order to consistently
calculate cloud formation and radiative transfer with their
feedback on convection and gas phase depletion.  We calculate the complete 1D
model atmosphere structure and the chemical details of the cloud
layers. The {{\sc Drift-Phoenix}} models enable the first stellar
atmosphere simulation that is based on the actual cloud {{\it
formation}} process. The resulting $(T,p)$ profiles differ
considerably from the previous limiting {{\sc Phoenix}} cases {{\sc
Dusty}} and {{\sc Cond}}. A tentative comparison with observations
demonstrates that the determination of effective temperatures based on
simple cloud models has to be applied with care.  Based on our new
models, we suggest a mean T$_{\rm eff}=1800$K for the L\,-\,dwarf
twin-binary system \objectname{DENIS J0205-1159} which is up to 500K
hotter than suggested in the literature. We show transition spectra
for gas-giant planets which form dust clouds in their atmospheres and
evaluate photometric fluxes for a \objectname{WASP-1} type system.
\end{abstract}

%% Keywords should appear after the \end{abstract} command. The uncommented
%% example has been keyed in ApJ style. See the instructions to authors
%% for the journal to which you are submitting your paper to determine
%% what keyword punctuation is appropriate.


%% From the front matter, we move on to the body of the paper.
%% In the first two sections, notice the use of the natbib \citep
%% and \citet commands to identify citations.  The citations are
%% tied to the reference list via symbolic KEYs. The KEY corresponds
%% to the KEY in the \bibitem in the reference list below. We have
%% chosen the first three characters of the first author's name plus
%% the last two numeral of the year of publication as our KEY for
%% each reference.

%% Authors who wish to have the most important objects in their paper
%% linked in the electronic edition to a data center may do so by tagging
%% their objects with \objectname{} or \object{}.  Each macro takes the
%% object name as its required argument. The optional, square-bracket 
%% argument should be used in cases where the data center identification
%% differs from what is to be printed in the paper.  The text appearing 
%% in curly braces is what will appear in print in the published paper. 
%% If the object name is recognized by the data centers, it will be linked
%% in the electronic edition to the object data available at the data centers  
%%
%% Note that for sources with brackets in their names, e.g. [WEG2004] 14h-090,
%% the brackets must be escaped with backslashes when used in the first
%% square-bracket argument, for instance, \object[\[WEG2004\] 14h-090]{90}).
%%  Otherwise, LaTeX will issue an error. 

\section{Introduction}
Today's most efficient tools to interpret the observed spectra of
substellar objects, i.e., brown dwarfs and planets, are 1D static
atmosphere simulations. Comparisons with observations, based on the
solution of the radiative-transfer problem which is done in great
detail with respect to the gas-phase opacities (Tsuji 2002, 2005;
Allard et al. 2001, Ackerman \& Marley 2001, Burrows \& Sharp 1999,
Barman et al. 2005), ideally yield insight into the atmospheric
structure and chemistry providing finger prints of its evolutionary
state.

 Brown dwarf and planetary atmospheres have a far more complex
chemistry than stellar objects due to the formation of clouds, which
bind chemical elements and, hence, strongly influences the remnant gas
phase inside the atmosphere. The presence of such clouds was
previously simplified in static model atmosphere codes.  We now move a
significant step forward by kinetically treating the chemistry of
cloud formation as a phase-non-equilibrium process in the framework of
model atmosphere simulations. Our dust model has been studied so far
for a given $(T, p, v_{\rm conv})$ structure ($T$ - gas temperature,
$p$ - gas pressure, $v_{\rm conv}$ - convective velocity), and we
present in this letter the first consistent simulation of cloud
micro-physics and atmospheric structure which allows for the first
time to study the feedback of the dust {\it formation} onto the
atmospheric structure. We present our first consistent {\sc
Drift-Phoenix} results together with a tentative comparison to the
observed spectrum of \objectname{ DENIS J0205--1159} and point out
possible uncertainties in present T$_{\rm eff}$ determinations
(Sect.~\ref{s:appl}).  We, however, leave aside the issue of
hydrodynamical cloud formation (e.g. Showman et al. 2006, Knutson et
al. 2007, Rauscher et al. 2007) which unavoidably has to deal with the
turbulent closure problem (Helling et al. 2004, Helling 2007). We
calculate transition spectra for gas-giant planets and a {\sc WASP}-1
type star, and evaluate the photometric fluxes for the 2MASS system,
for the VISIR system, and for the IRAC band systems.

\section{Method}\label{s:m}
We combined our model describing non-equilibrium dust formation ({\sc
 Drift}; Woitke \& Helling 2003, 2004; Helling \& Woitke 2006;
 Helling, Woitke, Thi 2008) with our general-purpose model atmosphere
 code ({\sc Phoenix}; Hauschildt \& Baron 1999). We consider a
 stationary dust formation process, where seeds form (homo-molecular
 TiO$_2$-nucleation, see Woitke \& Helling 2004, Sect.~2.2) from a
 highly supersaturated gas, grow to macroscopic particles of $\mu$m
 size, gravitationally settle into deeper layers, and eventually
 evaporate as the local temperature becomes too high for thermal
 stability (Woitke \& Helling 2003). Coagulation, a process amongst
 existing grains, is omitted so far since it acts on longer time and
 size scales than the dust formation processes considered here.
Typical time scales to form a 0.1$\mu$m SiO$_2$-grain in a gas of
 $T=1500$K, $\rho=10^{-6}$ g\,cm$^{-3}$ by continuous growth and
 coagulation are $\tau_{\rm gr}\approx10^{-1}$s (Fig. 3, Woitke \&
 Helling 2003) and $\tau_{\rm coag}\approx 10$\,s
 (Eq. 37[\footnote{All input quantities are taken from the model
 presented in this paper.}], Rossow 1978), respectively.  The
settling time scale is $\tau_{\rm grav}\approx 10^5$s (Eq. 20, Woitke
\& Helling 2003) for this grain crossing a pressure scale hight
($H_{\rm p}=10^6$cm) assuming a constant drift velocity $v_{\rm drift}\approx
10$cm\,s$^{-1}$. $v_{\rm drift}(a)$ rapidly increases for
$\rho>10^{-6}$ g\,cm$^{-3}$, hence  $\tau_{\rm grav}$ decreases inward.

A truly static atmosphere would not contain any dust (Woitke \&
 Helling 2004), therefore we include mixing by convection and
 overshooting by assuming an exponential decrease of the mass exchange
 frequency in the radiative zone (Eq. 9 in (Woitke \& Helling 2004)
 with $\beta=2.2$ and $\tau_{\rm mix}^{\rm min}=2/(H_{\rm p} v_{\rm conv})$), which serves
 to replenish the upper atmospheres and keeps the cycle of dust
 formation running. In order to keep the computing time reasonable, we
 consider the growth/evaporation of 7 solids (MgSiO$_3$[s],
 Mg$_2$SiO$_4$[s], MgO[s], SiO$_2$[s], SiO[s], Al$_2$O$_3$[s],
 TiO$_2$[s]) made of 5 different elements for which we solve 15
 possibly stiff conservation equations for 256 atmospheric layers. The
 whole 1D atmosphere problem is solved iteratively in {\sc
 Drift-Phoenix} where {\sc Phoenix} provides the actual $(T, p, v_{\rm
 conv})$ structure and {\sc Drift} solves the dust moment and element
 conservation equations in the subsonic, large Knudsen number
 case. {\sc Drift} hands back $(f(V, z), V_{\rm s}(z), \epsilon_{\rm i
 (i=Mg, Si, O, Al, Ti)})$ with $f(V, z)$ the parameterised
 distribution function of grain volume $V$ at atmospheric height $z$,
 $V_{\rm s}$ the chemical dust composition in volume fractions for all
 solids $s$ involved, and $\epsilon_{\rm i}$ the remaining element
 abundances for all involved elements $i$ in the gas phase (Dehn
 2007). Effective medium and Mie theory are used to calculate the
 opacity of these composite and chemically heterogeneous dust grains
 in {\sc Phoenix} (see Helling, Woitke \& Thi (2008), Table 2).

 \section{Results}\label{s:res}
 
\begin{figure}
\epsscale{1.0}
\plotone{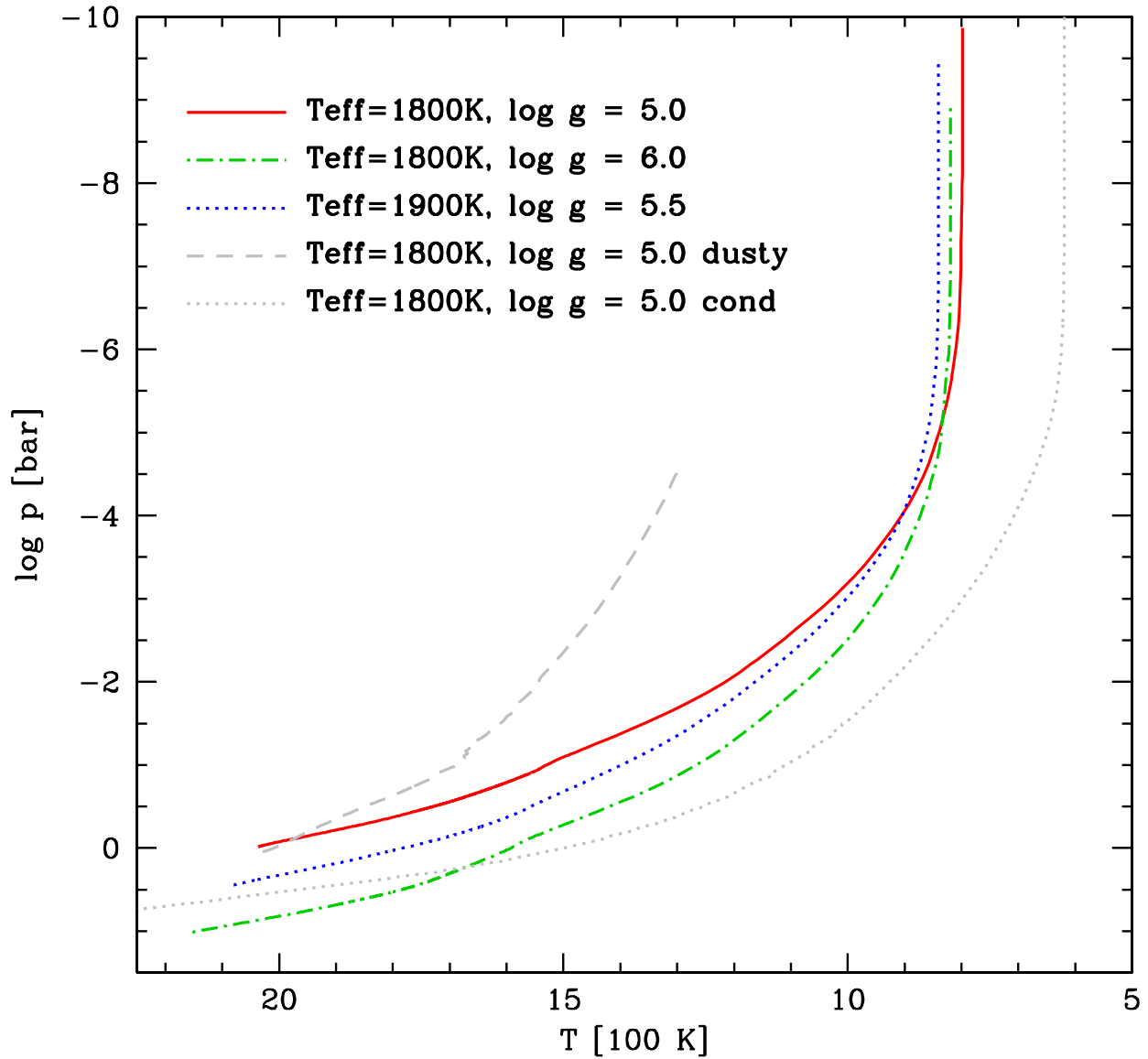}
%\plotone{TPStrucs.eps}
\caption{Temperature-pressure $(T, p)$ profiles for T$_{\rm
eff}=1800$K, $\log\,g=5.0$ (solar abundances) as result of different
dust modelling ({\sc Dirft}, {\sc Dusty}, {\sc Cond}) in the same model atmosphere code ({\sc Phoenix}). We show for comparison models with (T$_{\rm eff}=1800$K, $\log\,g=6.0$) and
(T$_{\rm eff}=1900$K, $\log\,g=5.5$); see also Fig.~\ref{f:Denis}.}
\label{f:Tp}
\end{figure}

 Figure~\ref{f:Tp} shows that {\sc Drift-Phoenix} $(T, p)$-profiles
 for late L-type brown dwarfs basically fall between the two limiting
 cases of the previous {\sc Phoenix} versions {\sc
 Dusty}\footnote{{\sc Dusty}: Dust is considered as opacity source and
 element sink.} and {\sc Cond}\footnote{{\sc Cond}: Dust is considered
 as element sink only.} (Allard et al. 2001). {\sc Dusty-Phoenix}
 models are hotter and {\sc Cond-Phoenix} models are cooler than our
 self-consistent model.  We notice that, amongst the (T$_{\rm
 eff}=1800$K, $\log\,g=5.0$) models, the {\sc Drift-Phoenix} profile
 has the highest temperatures inside the convectively active region
 below 1 bar.  For a given local temperature, the atmospheric pressure
 increases inside the dust forming region at $T > 1000K$ 
%(compare (T$_{\rm eff}=1800$K, $\log\,g=6.0$); Fig.~\ref{f:Tp}) 
with
 increasing surface gravity as suggested by previous models.  The
 result is a cloud being thermally stable at higher temperatures,
 hence sitting deeper in the atmosphere.

\begin{figure}
\epsscale{1.0}
\plotone{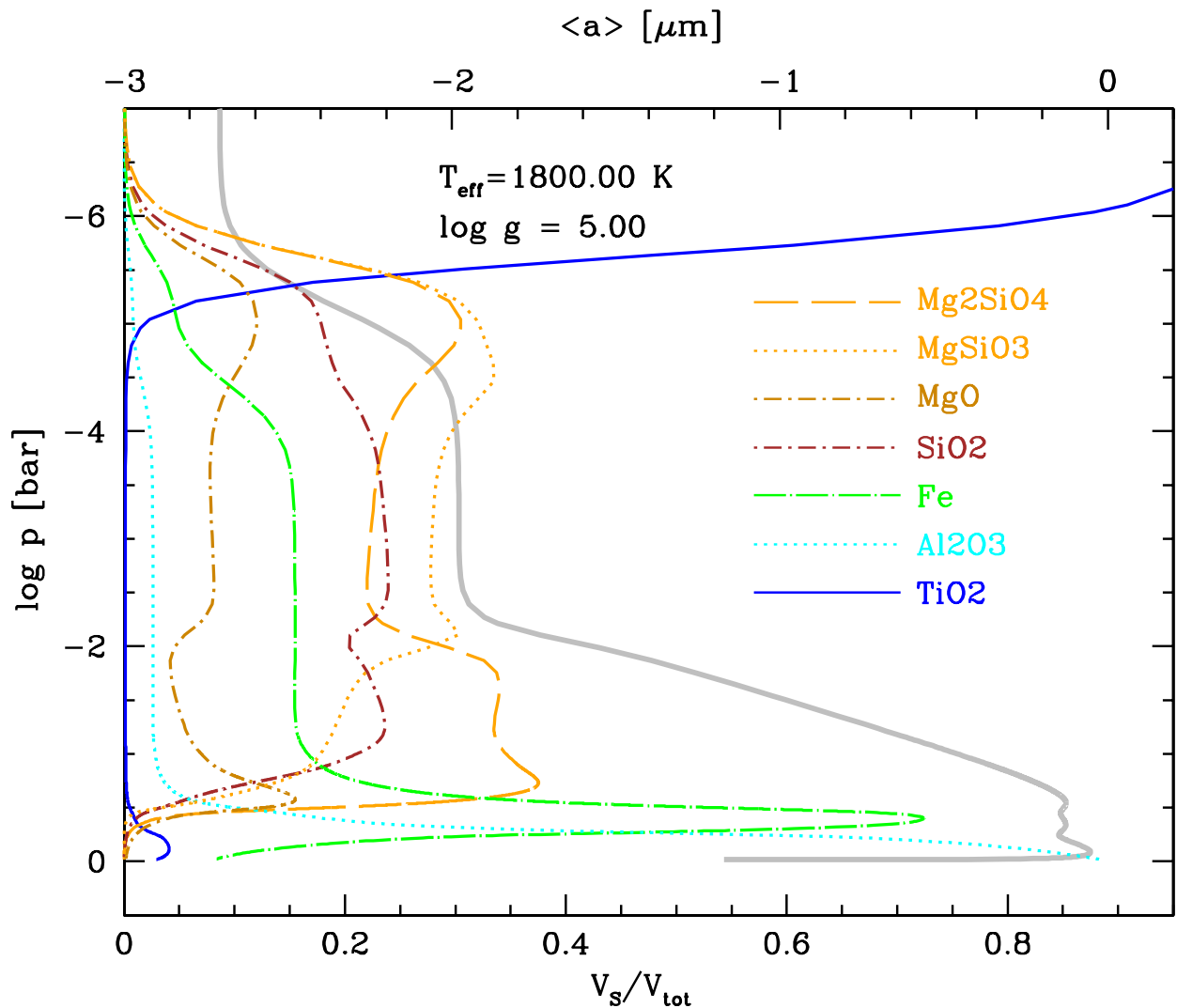}
%\plotone{Volfrac.eps}
\caption{Material composition of the dust cloud in volume fractions
$V_{\rm s}/V_{\rm tot}$ ($V_{\rm s}$ - volume fraction of solid s,
$V_{\rm tot}$ - total dust volume) and the mean particle
size $\langle a\rangle$ [$\mu$m] (thick gray) for the (T$_{\rm
eff}=1800$K, $\log\,g=5.0$) model with solar element abundances.}
\label{f:Vs}
\end{figure}

\begin{figure}
\epsscale{1.0}
\plotone{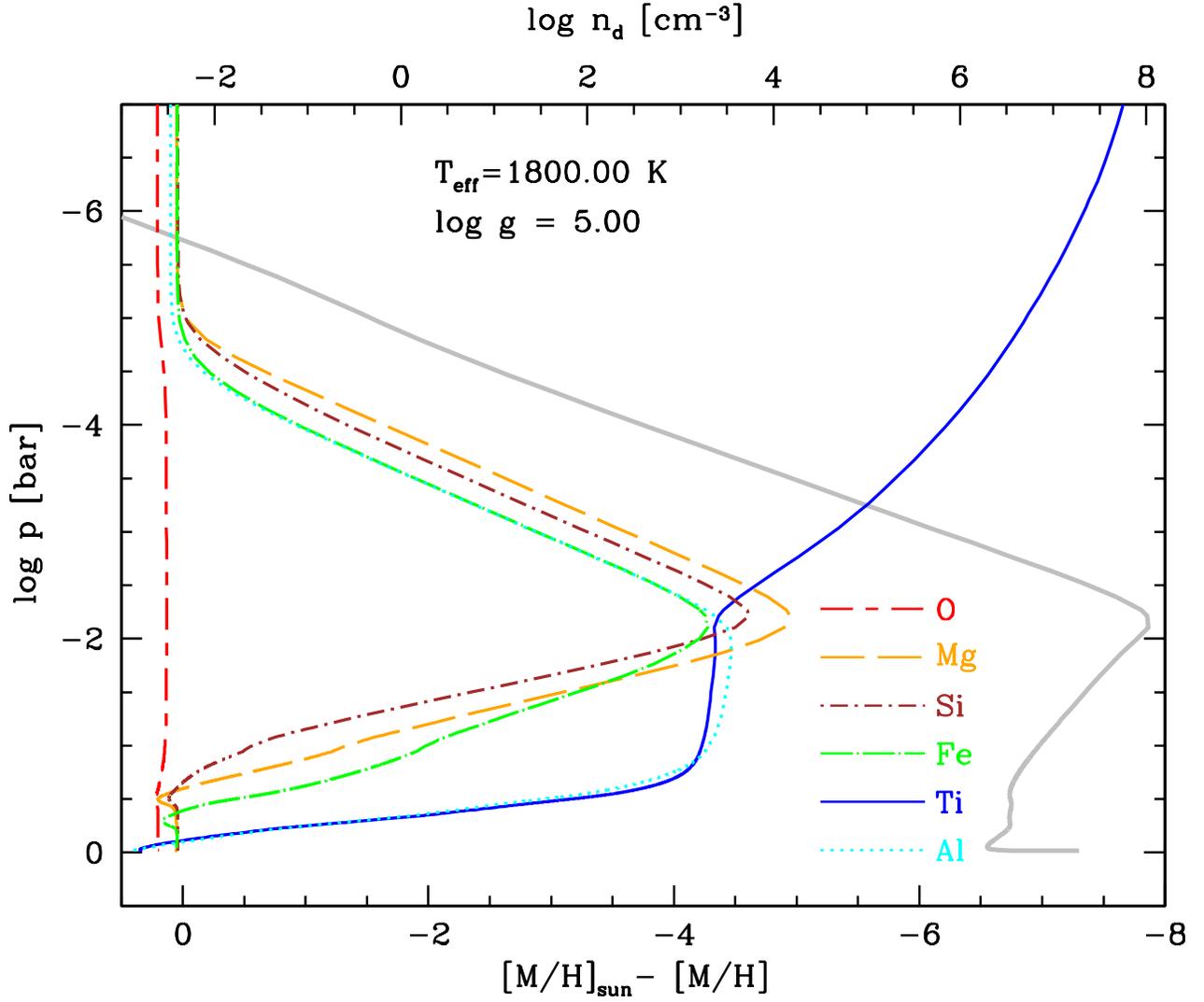}
%\plotone{Metalicity_nd.eps}
\caption{Metalicity $\log(\epsilon^{\rm 0}_{\rm i}/\epsilon_{\rm H}) -
\log(\epsilon_{\rm i}/\epsilon_{\rm H})$ ($\epsilon^{\rm 0}_{\rm i}$ -
solar abundances of element $i$, H - hydrogen) and dust number density
$n_{\rm d}$ [cm$^{-3}$] (thick gray) for the (T$_{\rm
eff}=1800$K, $\log\,g=5.0$) model.}
\label{f:eps}
\end{figure}

Grain sizes $\langle a\rangle$, material composition $V_{\rm s}/V_{\rm
 tot}$, number of dust particles $n_{\rm d}$ and remaining element
 abundances $\epsilon_{\rm i}$ determine the gas-phase opacity and are
 major ingredients for the radiative transfer
 calculation. Figure~\ref{f:Vs} shows for the model with (T$_{\rm
 eff}=1800$K, $\log\,g=5.0$) that the upper cloud layer is mainly made
 of small silicate grains with only some MgO[s] and Fe[s] impurities
 which is in accordance with our previous results (Helling et
 al. 2006, Helling, Woitke \& Thi 2008). However, only a few of such
 silicate grains populate the cloud deck (Fig.~\ref{f:eps} thick gray
 line). The material composition changes drastically at $\sim\,0.1$bar
 where big grains form the majority of the grain population.  The dust
 cloud formation causes a local element sink (Fig.~\ref{f:eps}).  The
 strongest element depletion coincides with the maximum number of
 particles in pressure space.  High-temperature condensable elements
 like Al and Ti are depleted in a wider atmosphere range since the
 associated solids are thermally stable over a larger temperature
 range than those binding Mg and Si.

\section{Application}\label{s:appl}

\paragraph{Brown Dwarf spectra:}
{\sc Drift-Phoenix} was used to tentatively reproduce the observed
spectrum of the L\,-\,dwarf twin-binary system {\sc Denis} J0205-1159
observed between 0.5$\mu$m and $2.6\,\mu$m (Reid et al. 2001; Leggett
et al. 2001).  The determination of its mean effective temperatures
proves difficult: Leggett et al. (2001) suggested T$_{\rm eff}=$1900K
and $\log\,g = 5.5$ based on {\sc Dusty-Phoenix} models and T$_{\rm
eff}=1400\,\ldots\,1600$K from structural models. Golimowsky et
al. (2004) suggest, depending on the system's age, an effective
temperature between 1350K and 1700K. Vrba et al. (2004) suggest
T$_{\rm eff}=$1563K with an astonishing precision.  Brandeker et
al. (2006) report similar difficulties in determining stellar
parameter of the binary OPH 162225-240515. Based on {\sc
Dusty-Phoenix} and Burrows models (Burrow et al. 2006) they suggest
T$_{\rm eff}=(2350 \pm 150)$K and T$_{\rm eff}=(2100\pm 100)$K for their
components A an B, respectively.

\begin{figure}
\epsscale{1.0}
\plotone{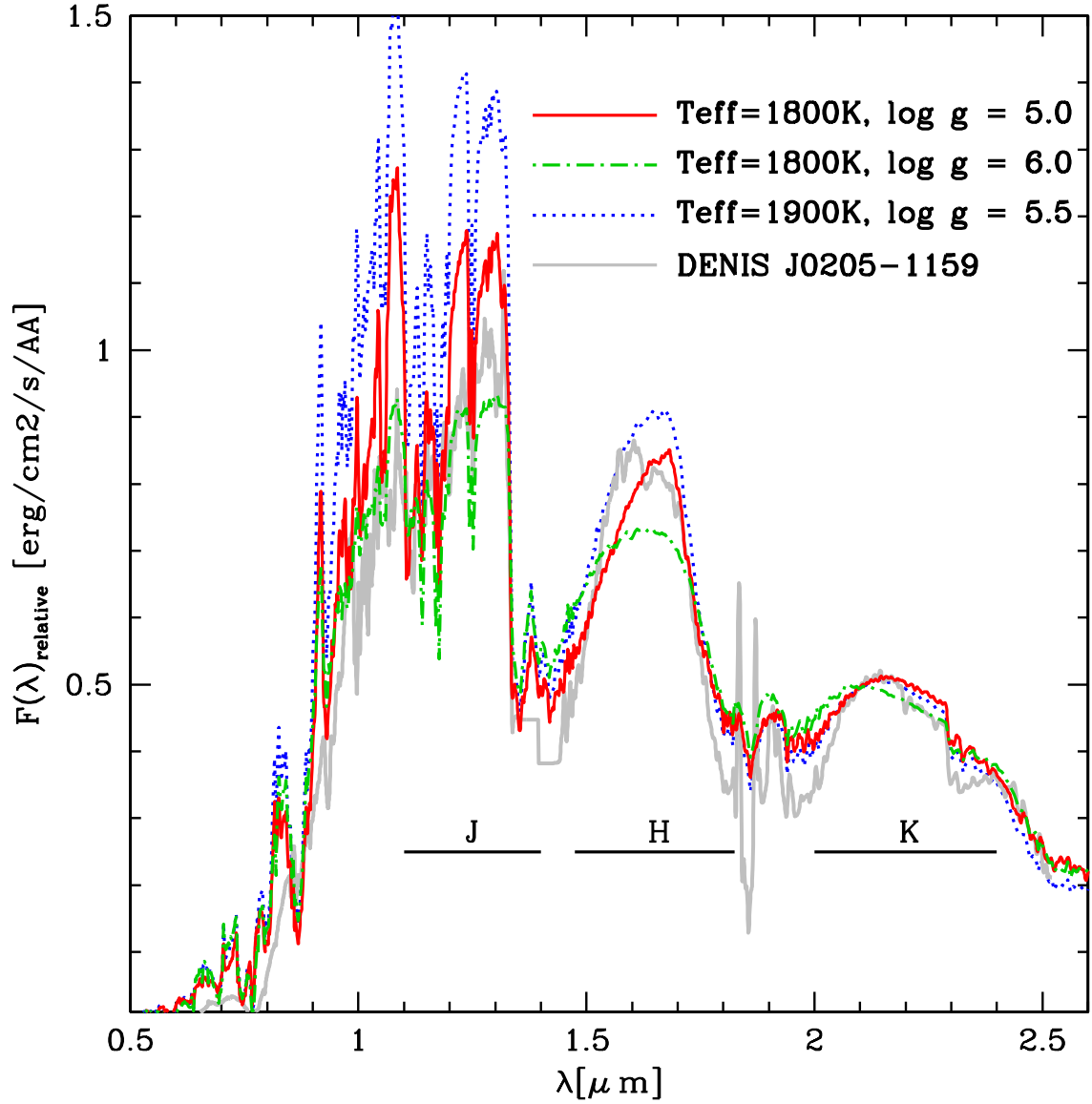}
%\plotone{SpecFitDENIS0205.eps}
\caption{{\sc Drift-Phoenix} synthetic spectra in comparison to the
brown dwarf twin-binary {\sc Denis} J0205-1159 spectrum from Reid et
al. (2001). The synthetic spectra are normalised to the
observed spectrum at $\lambda=2.1\mu$m, they are Gaussian convolved to
3400 sampling points with a spectral resolution of R=600. For $(T, p)$ structures see  Fig.~\ref{f:Tp}.}
\label{f:Denis}
\end{figure}

Figure~\ref{f:Denis} shows our comparison of {\sc Drift-Phoenix}
synthetic spectra of T$_{\rm eff}=$1800K with $\log\,g =5.0$ /
$\log\,g =6.0$ (red solid / green dash-dot line) and T$_{\rm
eff}=$1900K with $\log\,g=5.5$ (blue dotted) to the {\sc Denis}
J0205-1159 spectrum (gray) between 0.5$\mu$m and $2.6\,\mu$m. The
synthetic spectra are normalised to the observed spectrum in the
K-band at $\lambda=2.1\,\mu$m\,(\footnote{Cushing et al. (2007)
normalise to $1.25\mu$m in the J, $1.6\mu$m in the H, and $2.1\mu$m in
the K band.}).  The stellar parameters have been chosen to produce the
overall best fit in this spectral range but we cannot identify a
single model that fits the whole wavelength interval.  The J-band data
are best reproduced by the (T$_{\rm eff}=$1800K, $\log\,g=6.0$) model.
The H- and K-band data are best reproduced by the (T$_{\rm
eff}=$1800K, $\log\,g=5.0$) model. Note that the H band is most
sensitive to chances in the surface gravity of the models.  The
(T$_{\rm eff}=$1900K, $\log\,g=5.5$) model in general produces too
much flux except for the long-wavelength edge of the K band.
Tsuji (2005) and Cushing et al. (2007) report similar challenges for
their models in simultaneous fitting the JHK bands.  All shown
models reproduce very well the H$_2$O absorption edge longword
$1.3\,\mu$m and $1.75\,\mu$m.

We tentatively suggest an effective temperature of T$_{\rm
eff}=$1800K. Our tentative suggestion for the surface gravity is
$\log\,g =5.0$ since an increase in $\log\,g$ broadens the H band peak
and a decrease would narrow it, both causing even stronger
discrepancies to the observed spectrum.  It appears unlikely,
according to our model, that the system's mean temperature should be
as low as T$_{\rm eff}\lesssim1600$K, since the synthetic flux would
drop considerably.\\ Our results furthermore suggest that L\,-\,dwarfs
contain a considerable amount of dust in their atmospheres with a
gradually changing chemical dust composition (see
Fig.~\ref{f:Vs}). This explains also why our simulations yield about
the same T$_{\rm eff}$ range as those in Leggett et al. (2001) who
applied the {\sc Dusty-Phoenix} limiting case, where the dust was kept
in the atmosphere as opacity source, and consequently a high
atmospheric dust content was assumed. The actual $(T, p)$-profiles,
however, are very different with all its implications for the
gas-phase and cloud formation chemistry (see
Fig.~\ref{f:Tp}). Compared to T$_{\rm eff}=1350\,-\,1700$K (Golimowsky
et al. 2004), our results would then suggest an age of 2 Gyr.\\ So
far, our synthetic spectra are the results of a detailed
micro-physical treatment of dust formation in phase-non-equilibrium
including the formation of seed particles, their growth, evaporation,
and the effect of gravitational settling.  Grain sizes and grain size
distribution, cloud thicknesses, dust composition, total dust content
etc. are results of the simulations and are not prescribed as it was
needed in the {\sc Dusty-} and {\sc Cond-Phoenix} models.  Our
simulations, however, leave space for further improvement of our
synthetic spectra like for instance regarding the treatment of alkali
line profiles in the optical (Allard et al. 2007, Johnas et al. 2008),
the formation of liquid water droplets, or the improvement of our
mixing modelling.

%{\ }\\*[-1.2cm]

\begin{figure}
\epsscale{1.0}
\plotone{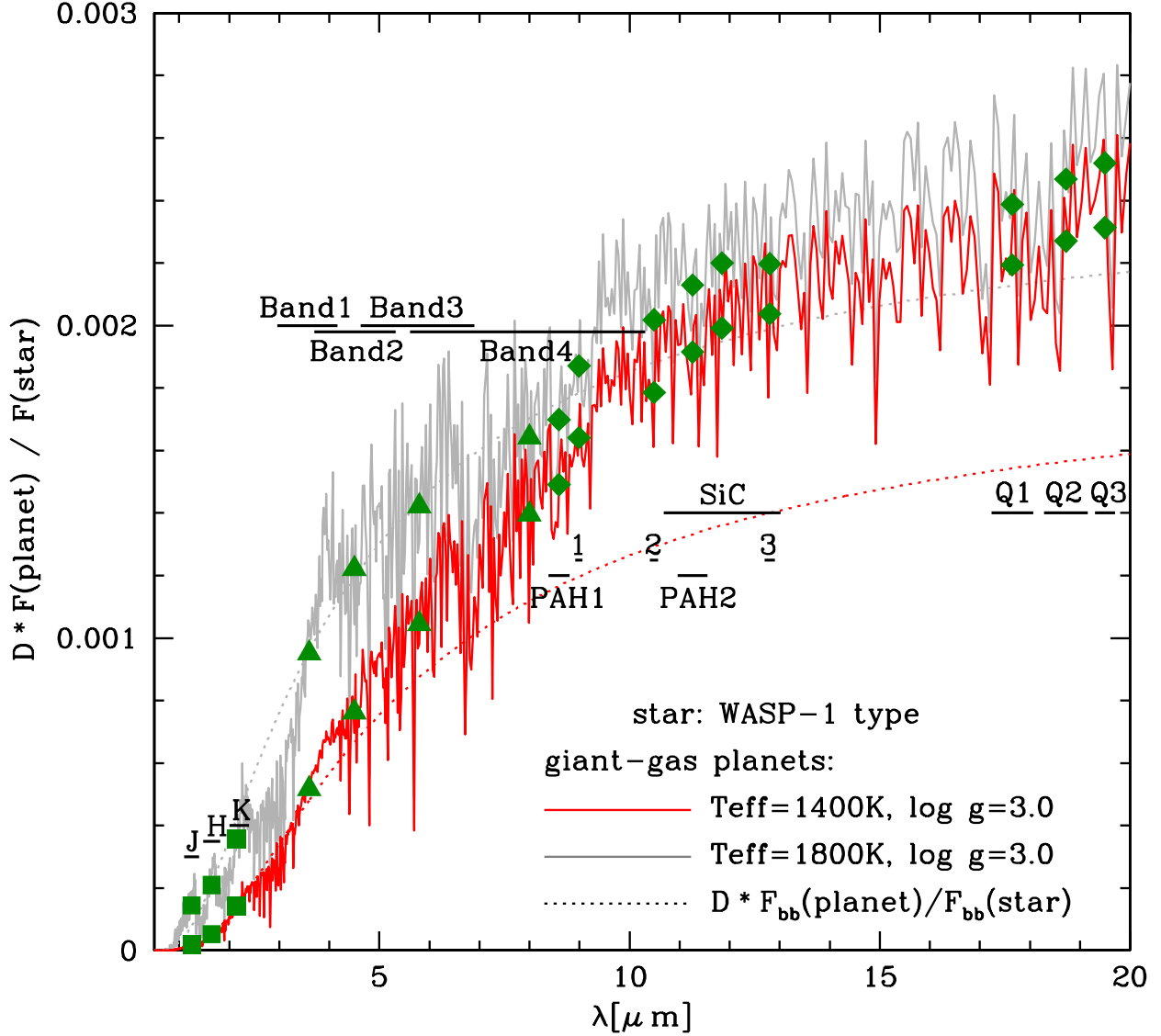}
%\plotone{GasGiantTransSpecs.eps}
\caption{Transition spectra for two gas-giant planets seen
 in the light of a WASP-1 type host star (T$_{\rm eff}=6140$K,
 $\log\,$g=4.31; compare Collier Cameron et al. 2007).  The synthetic
 spectra are Gauss convolved to 3000 sampling points with R=15000.
 Horizontal lines indicate the photometric bands and the symbols give
 photometric fluxes for the J, H, K bands ($\blacksquare$), the VISIR
 bands ($\blacklozenge$) PAH1, ArIII\,(\,1\,), SIV\,(\,2\,), PAH2, SiV\,(\,3\,), SiC, and the IRAC bands Band 1\,--\,4 ($\blacktriangle$).}
\label{f:transspec}
\end{figure}

\paragraph{Gas-giant transition spectra:}
 We have calculated transition spectra for a hot and a medium
 temperature gas-giant planet with T$_{\rm eff}=1800$K and T$_{\rm
 eff}=1400$K, respectively, and $\log\,g =3.0$ in the light of a WASP-1
 type host star (T$_{\rm eff}=6140$K, $\log\,g =4.31$, solar metallicity;
 compare Collier Cameron et al. 2007). We neglect, as a first attempt,
 the irradiation of the host star and day-night effects. For example,
 Barman et al. (2005) and Burrows et al. (2006) include these effects
 but do not account for the influence of cloud formation. 
 Figure~\ref{f:transspec} shows $D \times F_{\lambda}^{\rm
 planet}/F_{\lambda}^{\rm star}$, the transition spectra between
 $0.5\,\mu$m and 15$\,\mu$m corrected by the surface area ratio
 $D=(R_{\rm planet}/R_{\rm star})^2$ ($R_{\rm
 planet}=8\,10^9$cm$=1.11\,R_{\rm Jupiter}$, $R_{\rm
 star}=8.63\,10^{10}$cm$=1.24\,R_{\odot}$; $M_{\rm planet}=0.5\,M_{\rm
 Jupiter}$, $F_{\lambda}$ - surface flux). We add $D \times F_{\rm
 bb}(T_{\rm eff}^{\rm planet})/F_{\rm bb}(T_{\rm eff}^{\rm star})$ and
 the photometric fluxes for the planets thermal emission for the VISIR
 and IRAC bands, and the 2MASS J\,H\,Ks bands for comparison. The
 transit contrast due to the thermal emission increases with
 increasing wavelength but the difference between the T$_{\rm
 eff}=1800$K and T$_{\rm eff}=1400$K model become constant for
 $\lambda>8\mu$m.  Our cloud-covered gas-giant planets produce
 photometric contrasts which are comparable to the secondary eclipse
 measurements by Charbonneau et al. (2005), Deming et al. (2007),
 Snellen \& Covino (2007). We furthermore note a good agreement with
 the results on HD\,209458b by Barman et al. (2005) despite all the
 uncertainties in the cloud and the day-night-effect modelling.
 
\section{Summary}

The modelling of cloud formation in substellar atmospheres has a
strong impact on the objects temperature-pressure structure which in
turn determines the cloud's chemical composition, the grain size
distribution function, and the cloud's location inside the
atmosphere. These are results of our consistent simulation of
substellar atmospheres and detailed non-equilibrium dust cloud
formation.  We demonstrate synthetic transition spectra for gas-giant
planets and calculate synthetic photometric fluxes based on our
consistent solution of dust cloud formation and radiative transfer
problem.  A future goal is to extend our model to solar-system-like
planets with much cooler atmosphere.

As we have shown for the field object {\sc Denis} J0205-1159, stellar
parameter determinations based on a comparison with synthetic spectra
can vary considerable. As one consequence of this, we have set out to
conduct a component-based study
\footnote{http://phoenix.hs.uni-hamburg.de/BrownDwarfsToPlanets1/ }
where e.g. cloud compositions, dust-to-gas-ratios, and grain sizes are
compared for different cloud models (Helling et al. 2007, 2008).

\acknowledgments We thank Aleks Scholz, Andrew Collier Cameron and the
anonymous referee for helpful discussions on the paper's subject.  The
computer support at the School of Physics and Astronomy St Andrews is
highly acknowledged.  Most of the literature search was done with ADS.

\clearpage

%% Use the figure environment and \plotone or \plottwo to include
%% figures and captions in your electronic submission.
%% To embed the sample graphics in
%% the file, uncomment the \plotone, \plottwo, and
%% \includegraphics commands
%%
%% If you need a layout that cannot be achieved with \plotone or
%% \plottwo, you can invoke the graphicx package directly with the
%% \includegraphics command or use \plotfiddle. For more information,
%% please see the tutorial on "Using Electronic Art with AASTeX" in the
%% documentation section at the AASTeX Web site,
%% http://www.journals.uchicago.edu/AAS/AASTeX.
%%
%% The examples below also include sample markup for submission of
%% supplemental electronic materials. As always, be sure to check
%% the instructions to authors for the journal you are submitting to
%% for specific submissions guidelines as they vary from
%% journal to journal.

%% This example uses \plotone to include an EPS file scaled to
%% 80% of its natural size with \epsscale. Its caption
%% has been written to indicate that additional figure parts will be
%% available in the electronic journal.

\clearpage

%% Here we use \plottwo to present two versions of the same figure,
%% one in black and white for print the other in RGB color
%% for online presentation. Note that the caption indicates
%% that a color version of the figure will be available online.
%%

\clearpage

\end{document}